\documentclass[prb,11pt,notitlepage]{revtex4-2}

\pdfoutput=1

\usepackage{amsmath}  
\usepackage{graphicx} 
\usepackage{verbatim} 
\usepackage{color}    
\usepackage{subcaption} 
\usepackage{hyperref}  
\usepackage{float}
\usepackage{physics}
\usepackage{verbatim}

\DeclareMathOperator{\sn}{sn}
\DeclareMathOperator{\cn}{cn}
\DeclareMathOperator{\dn}{dn}
\DeclareMathOperator{\E}{E}

\begin{document}
\author{Christian Peterson}
\email{cpeter16@uccs.edu}
\affiliation{Dept. of Physics, University of Colorado at Colorado Springs, Colorado Springs, CO, 80918}
\date{11/21/2022}
\begin{abstract}
Exact solutions are found for Euler's equations of rigid body motion for general asymmetrical bodies under the influence of torque by using Jacobi elliptic functions. Differential equations are determined for the amplitudes and the parameters of the elliptic functions. The solution is then applied to the detumbling of a satellite with arbitrary initial rotation rates where numerical solutions are seen to be in agreement with the analytical solution. The body fixed frame solution is then transformed to the inertial frame by use of a quaternion rotation matrix to depict the motion in figures and in animations within a Mathematica notebook which is openly published on the Wolfram community.
\end{abstract}
\title{Exact solutions to Euler's equations for rigid body motion with application to detumbling satellites}
\maketitle
\section{Introduction}
The rotational motion about the center of mass in the coordinate frame that is fixed to the body is described by Euler's equations of motion. The system of three ordinary differential equations is coupled and non-linear, determining the dynamics of the angular velocities in the direction of the principal axes of the body. Rigid body motion has attracted the attention of many investigations due to its practical applicability to the attitude control of space vehicles and aircraft. Several studies have investigated analytical solutions to special cases of Euler's equations under torque, such as by assuming symmetric or near-symmetric bodies or by assuming one of the angular velocities is near zero\cite{Longuski1,Oldenburg,McNair,coppola}. A first-order approximation of a general rigid body subjected to torques is provided by Longuski and Tsiotras \cite{tsiotras}. Panayotounakos et al.\cite{panayotounakos} present a complete analytical solution for an asymmetric body by reducing Euler's equations to Abel differential equations of the second kind of the normal form. 

In this paper, an analytical solution to Euler's equations is presented by assuming a Jacobi elliptic function form for the angular velocities. The solution to Euler's equations for torque free motion is known to have Jacobi elliptic function solutions where the eccentricity and the amplitudes of oscillation are constant\cite{LL,Whittaker,peterson}. In the current investigation the parameters of the Jacobi elliptic functions are all assumed to have time dependence. Differential equations for the elliptic function parameters are derived by using the method of undetermined coefficients, which are then solved exactly by implementing special forms of torque. The analytical solution is shown to be in agreement to the numerical integration of Euler's equations. The solution is then applied to detumbling a satellite which is initially rotating with arbitrary initial conditions about its principal axes. It is transformed from the body-fixed frame to the inertial frame via a quaternion comprised rotation matrix. The principal axes, angular momentum, and torque are then presented to depict the evolution of the rotating satellite.
\section{Rigid Body Dynamics in the Body-Fixed Frame}
The Euler equations for rotational motion in the body fixed frame are,
\noindent 
\begin{align}
\dot\omega_1 &= \frac{(I_2 - I_3)}{I_1}\omega_2\omega_3 + \frac{\tau_1}{I_1},\\
\dot\omega_2 &= \frac{(I_3 - I_1)}{I_2}\omega_3\omega_1 + \frac{\tau_2}{I_2},\\
\dot\omega_3 &= \frac{(I_1 - I_2)}{I_3}\omega_1\omega_2 + \frac{\tau_3}{I_3}.
\end{align}
In this coordinate system, the moment of inertia tensor is diagonal with the constant principal moments $I_1$, $I_2$, $I_3$ and where the angular velocities along each respective axis are denoted as $\omega_1$, $\omega_2$, and $\omega_3$. The aim of the present analysis is to utilize Jacobi elliptic functions to solve for the time-dependent angular velocities in the body-fixed frame of an asymmetrical rigid body subjected to time varying torques $\tau_1$, $\tau_2$, and $\tau_3$.
One can arrange the moments of inertia without loss of generality such that,
\[I_1<I_2<I_3.\]
The following substitutions are made for positive quantity reduced moment coefficients,
\begin{align*}
    \mu_1 &= \frac{I_3-I_2}{I_1},\\ 
    \mu_2 &=\frac{I_3-I_1}{I_2},\\
    \mu_3 &=\frac{I_2-I_1}{I_3},
\end{align*} 
such that the euler equations may be re-written as,
\begin{align}
\dot\omega_1 &= -\mu_1\omega_2\omega_3 + \frac{\tau_1}{I_1},\\
\dot\omega_2& = \mu_2\omega_3\omega_1 + \frac{\tau_2}{I_2},\\
\dot\omega_3 &= -\mu_3\omega_1\omega_2 + \frac{\tau_3}{I_3}.  
\end{align}
The Jacobi elliptic functions $\sn(u, k)$, $\cn(u, k)$, and $\dn(u, k)$ can be defined like trigonometric functions using ratios on an ellipse where the argument $u$ is analogous to the angle variable, and a new argument $k$ denotes the eccentricity of the ellipse taking any value $0 \leq k \leq 1$. The periods of $\sn(u, k)$ and $\cn(u, k)$ are $4K(k)$, and the period of $\dn(u, k)$ is $2K(k)$, where $K(k)$ is the complete elliptic integral the first kind. As the eccentricity, $k$, of the reference ellipse defining the elliptic functions tends to zero $K(k)$ approaches $\pi/2$, and the functions $\sn(u,k)$ and $\cn(u,k)$ become  sine and cosine respectively while $\dn(u,k)$ becomes equal to one in the limit. The derivatives of the elliptic functions with respect to $u$ are,
\begin{align}
\dfrac{d}{du}\sn(u,k) &= \cn(u,k) \dn(u,k),\\
\dfrac{d}{du}\cn(u,k) &= - \sn(u,k) \dn(u,k),\\
\dfrac{d}{du}\dn(u,k) &= - k^2 \sn(u,k) \cn(u,k),
\end{align}
and derivatives with respect to modulus, in accordance with Dixon's \cite{Alfred} convention, are,
\begin{align}
\dfrac{d}{dk}\sn(u,k) &= \frac{k}{k'^2}\sn(u,k) \cn^2(u,k)+ \frac{u}{k}\cn(u,k)\dn(u,k)-\frac{\E(u,k)}{k k'^2}\cn(u,k)\dn(u,k).\\
\dfrac{d}{dk}\cn(u,k) &= -\frac{k}{k'^2}\sn^2(u,k) \cn(u,k)- \frac{u}{k}\sn(u,k)\dn(u,k)+\frac{\E(u,k)}{k k'^2}\sn(u,k)\dn(u,k),\\
\dfrac{d}{dk}\dn(u,k) &= -\frac{k}{k'^2}\sn^2(u,k) \dn(u,k)- k u \sn(u,k)\dn(u,k)+\frac{k \E(u,k)}{k'^2}\sn(u,k)\cn(u,k),
\end{align}
where $\E(u,k)$ is the incomplete elliptic integral of the second kind and $k'^2 = 1-k^2$. Assuming an elliptic function solution we set,
\begin{align}
    \omega_1 &= A_1(t) \dn(u(t),k(t)),\\    
    \omega_2 &= A_2(t) \sn(u(t),k(t)),\\
    \omega_3 &= A_3(t) \cn(u(t),k(t)),
\end{align}
where the angular velocity amplitudes $A_i(t)$ along with $u(t)$ and $k(t)$ all in general may vary with time. Another solution exists by exchanging the $\cn(u,k)$ and $\dn(u,k)$ functions, and the suitable selection depends on the ratios of $A_3(t)$ and $A_1(t)$. Taking the time derivatives of the assumed solution we have,
\begin{align}
    \dot\omega_1 = -\frac{A_1 k}{A_3 A_2}\left(\frac{k'^2(k\dot u + \dot k u)-E \dot k}{k'^2}\right)\omega_3\omega_2 + \dot A_1 \dn(u,k) - \frac{A_1k\dot k}{k'^2}\sn^2(u,k)\dn(u,k),\\
    \dot\omega_2 = \frac{A_2}{A_1 A_3}\left(\frac{k'^2(k\dot u + \dot k u)-E\dot k}{k k'^2}\right)\omega_1\omega_3 + \dot A_2 \sn(u,k) + \frac{A_2 k\dot k}{k'^2}\sn(u,k)\cn^2(u,k),\\
    \dot\omega_3 = -\frac{A_3}{A_2 A_1}\left(\frac{k'^2(k\dot u + \dot k u)-E\dot k}{k k'^2}\right)\omega_2\omega_1 + \dot A_3 \cn(u,k) - \frac{A_3 k\dot k}{k'^2}\sn^2(u,k)\cn(u,k).
\end{align}
Then by equating coefficients in equations (16), (17), and (18) to those in equations (4), (5), and (6) respectively we arrive at the following six equations,
\begin{align}
    \mu_1 &= \frac{A_1 k}{A_3 A_2}\left(\frac{k'^2(k\dot u + \dot k u)-E \dot k}{k'^2}\right),\\
    \mu_2 &= \frac{A_2}{A_1 A_3}\left(\frac{k'^2(k\dot u + \dot k u)-E\dot k}{k k'^2}\right),\\
    \mu_3 &= \frac{A_3}{A_2 A_1}\left(\frac{k'^2(k\dot u + \dot k u)-E\dot k}{k k'^2}\right),\\    
     \frac{\tau_1}{I_1} &= \dot A_1 \dn(u,k) - \frac{A_1k\dot k}{k'^2}\sn^2(u,k)\dn(u,k),\\
    \frac{\tau_2}{I_2} &= \dot A_2 \sn(u,k) + \frac{A_2 k \dot k}{k'^2}\sn(u,k)\cn^2(u,k),\\
    \frac{\tau_3}{I_1} &= \dot A_3 \cn(u,k) - \frac{A_3 k \dot k}{k'^2}\sn^2(u,k)\cn(u,k).
\end{align}
Multiplying Eq.(21) by the reciprocal of Eq.(20) we find that,
\begin{align}
    A_2(t) = \sqrt{\frac{\mu_2}{\mu_3}}A_3(t).
\end{align}
Similarly by dividing Eq.(19) by Eq.(21) it is determined that $k(t)$ is given by,
\begin{align}
    k(t) = \sqrt{\frac{\mu_1}{\mu_3}}\frac{A_3(t)}{A_1(t)}.
\end{align}
This solution is suitable so long as $k(t) < 1$. In the event where $A_3(t)$ becomes larger than $\sqrt{\frac{\mu_3}{\mu_1}}A_1(t)$ then the $\cn(u,k)$ and $\dn(u,k)$ functions exchange in the solution, and indices 1 and 3 interchange throughout the derivation. Any of the equations (19), (20), or (21) in conjunction with Eq.(25) provide the differential equation for $u(t)$,
\begin{align}
    \dot u(t) = \sqrt{\mu_3 \mu_2} A_1(t) + \frac{\dot k}{k}\left(\frac{\E(u,k)}{k'^2} - u(t)\right),
\end{align}
The differential equation for $A_3(t)$ is found from equations (23), (24), and (25) to be,
\begin{align}
    \dot A_3(t) = \frac{\tau_3}{I_3}\cn(u,k) + \sqrt{\frac{\mu_3}{\mu_2}}\frac{\tau_2}{I_2}\sn(u,k).
\end{align}
By noting that,
\begin{align*}
    \dot k(t) = k(t)\left(\frac{\dot A_3(t)}{A_3(t)}-\frac{\dot A_1(t)}{A_1(t)}\right),
\end{align*}
and using the identity,
\begin{align*}
    \sn^2(u,k) + \cn^2(u,k) = 1
\end{align*}
we use Eq.(22) to find the differential equation for $A_1(t)$,
\begin{align}
    \dot A_1(t) = \frac{\tau_1 k'^2}{I_3 \dn(u,k) (1-k^2\cn^2(u,k))} +  \dot A_3(t) \sqrt{\frac{\mu_1}{\mu_3}} \frac{k \sn^2(u,k)}{1-k^2\cn^2(u,k)}.
\end{align}
To solve Eq.(28) and Eq.(29) we assume torques of the following form,
\begin{align}
    \tau_1 &= \frac{f_1(t) I_1 \dn(u,k)(1-k^2\cn^2(u,k))}{k'^2}-f_3(t)\sqrt{\frac{\mu_1}{\mu_3}}\frac{I_1 \dn(u,k) k\sn^2(u,k)}{k'^2},\\
    \tau_2 &= f_3(t)\sqrt{\frac{\mu_2}{\mu_3}}I_2\sn(u,k),\\
    \tau_3 &= f_3(t) I_3 \cn(u,k),
\end{align}
where $f_1(t)$ and $f_3(t)$ are any integrable functions. Thus we have,
\begin{align}
    A_1(t) = \int_{0}^{t}f_1(t)dt,\\
    A_3(t) = \int_{0}^{t}f_2(t)dt.
\end{align}
If $f_1(t)$ and $f_3(t)$ are chosen such that $\dot k = 0$ then Eq.(27) reduces to a separable equation yielding,
\begin{align}
   u(t) =  \sqrt{\mu_3 \mu_2}\int_{0}^{t} A_1(t) dt,
\end{align}
and Eq. (22) simplifies such that the torque around the 1-axis is given by,
\begin{align}
    \tau_1 = f_1(t) I_1 \dn(u,k)
\end{align}
In this manner we may choose a desired solution to Euler's equations then fully specify the fixed-frame torques necessary to achieve the solution. In the situation where all of the torques are zero then $A_1$, $A_2$, and $k$ are constant while $u(t)$ is linear, which corresponds with the known torque free solution of rigid body motion.
\section{Application of Detumbling a Satellite}
The analytical solution provided in section II is applied to detumbling the satellite depicted in Figure 1 from initial angular velocities about all three principal axes. A link to the Mathematica
notebook on the Wolfram Cloud with the simulation is provided at the end of this section. The principal moments of inertia for the satellite are,
 \begin{align*}
     I_1 = 0.359903\ \text{kg}\cdot\text{m\textsuperscript{2}},\\
     I_2 = 0.462824\ \text{kg}\cdot\text{m\textsuperscript{2}},\\
     I_3 = 0.549196\ \text{kg}\cdot\text{m\textsuperscript{2}}.
 \end{align*}
They are calculated by using the moments of inertia of rectangular prisms and a cylinder in conjunction with Steiner's parallel axis theorem.
\begin{figure}[H]
  \centering
  \fbox{\includegraphics[width=.5\textwidth]{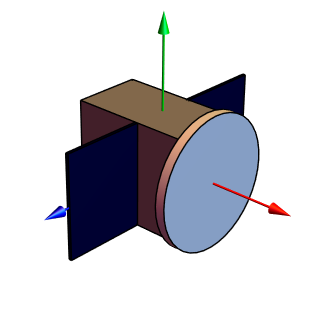}}     
  \caption{Satellite Model. The red, green, and blue arrows correspond respectively to the axes with I1, I2, I3 principal moments of inertia.}
\end{figure}
\noindent We choose Amplitude functions $A_1(t)$ and $A_3(t)$ to have a Gaussian exponential decay,
 \begin{align*}
     A_1(t) = A_{10}\ e^{-z t^2},\\
     A_3(t) = A_{30}\ e^{-z t^2},
 \end{align*}
 and the initial conditions applied to the satellite are chosen to be $A_{30} = 1$, $k = .35$, and $A_{10} = \sqrt{\frac{\mu_1}{\mu_3}}\frac{A_{30}}{k}$. The angular velocities given by equations (13), (14), and (15) are presented in figures 2, 3, and 4, depicted by the red line in each plot. The black dashed line in the figures are the plots of the numerical solution to Euler's equations provided by the torques given in equations (31), (32), and (36). The torques given are plotted in figures 5, 6 and 7.
 \begin{figure}[H]
    \centering
    \includegraphics[width=.5\textwidth]{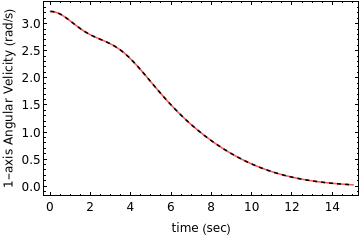}   
    \caption{Angular Velocity along the 1-axis. The red line is the Analytical solution and the black dashed line is the numerical solution to Euler's equations}
\end{figure}
\begin{figure}[H]
    \centering
    \includegraphics[width=.5\textwidth]{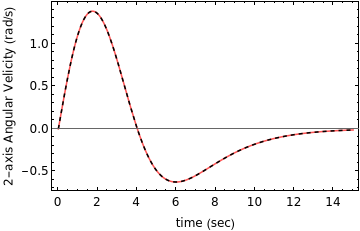}   
    \caption{Angular Velocity along the 2-axis. The red line is the Analytical solution and the black dashed line is the numerical solution to Euler's equations}
\end{figure}
\begin{figure}[H]
    \centering
    \includegraphics[width=.5\textwidth]{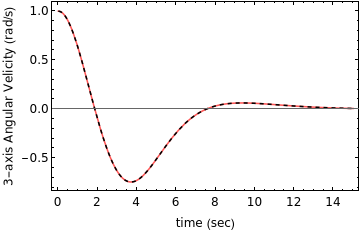}
    \caption{Angular Velocity along the 3-axis. The red line is the Analytical solution and the black dashed line is the numerical solution to Euler's equations}
\end{figure}
\begin{figure}[H]
    \centering
    \includegraphics[width=.5\textwidth]{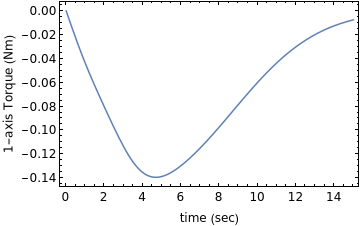}
    \caption{Torque along the 1-axis.}
\end{figure}
\begin{figure}[H]
    \centering
    \includegraphics[width=.5\textwidth]{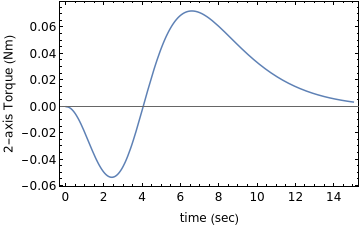}
    \caption{Torque along the 2-axis.}
\end{figure}
\begin{figure}[H]
    \centering
    \includegraphics[width=.5\textwidth]{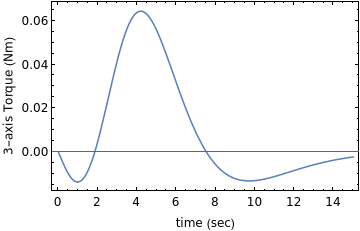}
    \caption{Torque along the 3-axis.}
\end{figure}
The solution of the body-fixed frame is transformed to the inertial frame using quaternions as discussed in detail by Coutsias, et al\cite{quaternions}. The quaternions $q_0$, $q_1$, $q_2$, and $q_3$ are solved for numerically by the equations,
\begin{align}
        \frac{d\Vec{q}}{dt}= \frac{1}{2}\left(\begin{array}{cccc}
              0 & -\omega_1 & -\omega_2 & -\omega_3 \\ 
              \omega_1 & 0 & \omega_3 & -\omega_2\\
              \omega_2 & -\omega_3 & 0 & \omega_1\\
              \omega_3 & \omega_2 & -\omega_1 & 0
           \end{array}\right)
           \left(\begin{array}{cccc}
              q_0\\ 
              q_1\\
              q_2\\
              q_3
           \end{array}\right).
\end{align}
The rotation matrix, Q, is then given by the quaternians by,
\begin{align}
        Q = 2 \left(\begin{array}{cccc}
              \frac{1}{2}(q_0^2 + q_1^2 - q_2^2 - q_3^2) & q_1q_2 - q_0q_3 & q_1q_3 + q_0q_2\\
              q_1q_2 + q_0q_3 & \frac{1}{2}(q_0^2 - q_1^2 + q_2^2 - q_3^2) & q_2q_3 - q_0q_1\\
              q_1q_3 - q_0q_2 & q_2q_3 + q_0q_1 & \frac{1}{2}(q_0^2 - q_1^2 - q_2^2 + q_3^2)\\
           \end{array}\right).
\end{align}
A vector $\Vec{x}$ in the body-fixed frame is transformed into the inertial frame by $\Vec{X} = Q^T \cdot \Vec{x}$. The rotation matrix is used to transform the principal axes, torque vector, and angular momentum vector into the inertial frame depicted in figures 8, 9 and 10. The torque and angular momentum vectors are normalized for the purpose of the graphics and are represented by the orange and black arrows respectively.
\begin{figure}[H]
    \centering
    \fbox{\includegraphics[width=.5\textwidth]{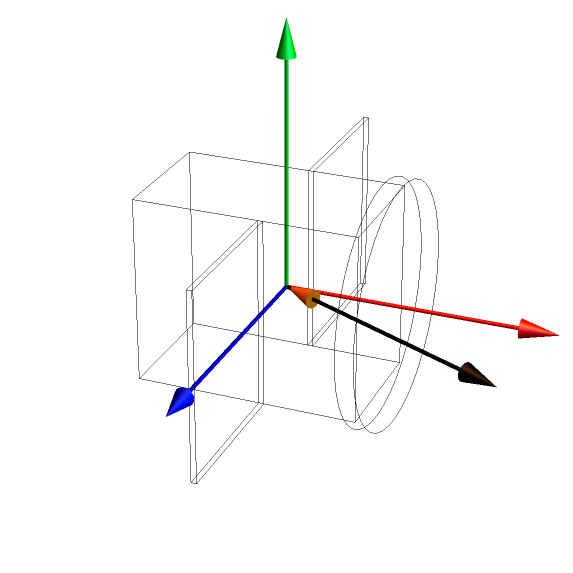}}
    \caption{Detumbling Satellite at t = 0s}
\end{figure}
\begin{figure}[H]
    \centering
    \fbox{\includegraphics[width=.5\textwidth]{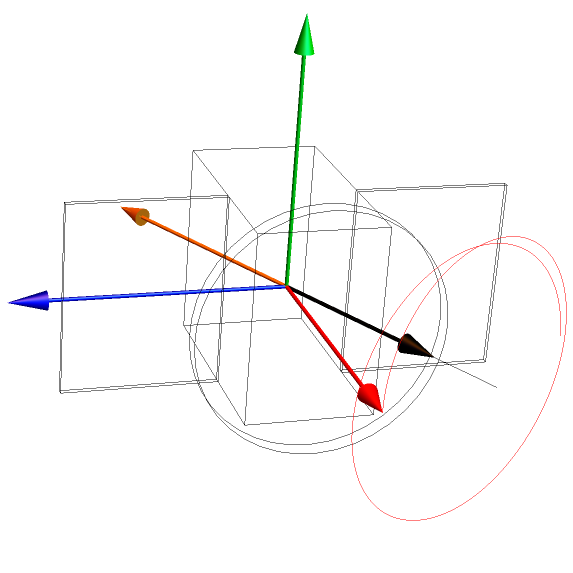}}
    \caption{Detumbling Satellite at t = 4s}
\end{figure}
\begin{figure}[H]
    \centering
    \fbox{\includegraphics[width=.5\textwidth]{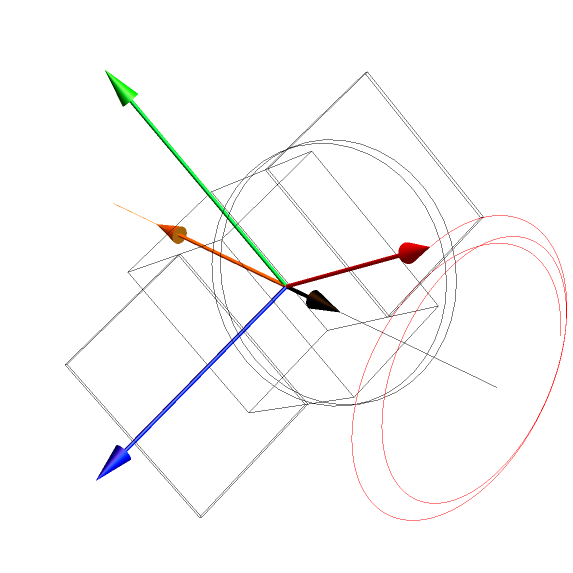}}
    \caption{Detumbling Satellite at t = 8s}
\end{figure}
The Mathimatica notebook used to plot and animate the solutions is published openly on the Wolfram community. It can be found at the following url.
\url{https://community.wolfram.com/groups/-/m/t/2701672}
\section{Discussion}
In this paper Jacobi elliptic functions are used to re-formulate and present analytical solutions to Euler's equations with torque. Equations (27), (28), and (29) for $u(t)$, $A_1(t)$ and $A_3(t)$ do not appear to be integrable for general torques, however by employing torques of the forms presented in equations (30)-(32), $A_1(t)$ and $A_3$ may be determined. We then solved for $u(t)$ by restricting the modulus to be constant.The solution is applied to detumbling a satellite, which may be useful in designing more efficient attitude control systems in spacecraft. Analysis using Mathematica's NDSolve function has shown that numerical solutions of the equations for $u(t)$, $A_1(t)$, and $A_3(t)$ are in agreement with  numerical solutions of Euler's Equations when the modulus is small. However, the solutions diverge as the modulus becomes larger, which provides an avenue for further investigation of the Jacobi elliptic function solutions.
\bibliographystyle{aipnum4-1}
\bibliography{References.bib}

\end{document}